\documentclass[12pt]{article}
\usepackage[english]{babel}
\usepackage{amsfonts}
\usepackage{amsmath}
\usepackage{amssymb}
\usepackage{mathrsfs}
\usepackage{multicol}
\usepackage{fancybox}
\usepackage{graphicx}
\usepackage{dsfont}
\usepackage{color}
\usepackage{hyperref}
\usepackage{graphicx} 

\def\snJ{\textrm{sn}}
\def\cnJ{\textrm{cn}}
\def\dnJ{\textrm{dn}}
\def\tl{\tilde} 
 
\def\gm{\gamma} 
\def\lm{\lambda}

\def\be{\begin{equation}}
\def\ee{\end{equation}}
\def\bea{\begin{eqnarray}}
\def\eea{\end{eqnarray}}

\def\dblone{\hbox{$1\hskip -1.2pt\vrule depth 0pt height 1.6ex width 0.7pt \vrule depth 0pt height 0.3pt width 0.12em$}}

\begin{document}

 \bigskip

\begin{center}
{\Large{\bf{B\"acklund transformations for the elliptic Gaudin model and a Clebsch system}}}
\end{center}
\bigskip

\begin{center}
{ {\bf Federico Zullo}}

{Dipartimento di Fisica,   Universit\`a di Roma Tre \\ Istituto Nazionale di
Fisica Nucleare, sezione di Roma Tre\\  Via Vasca Navale 84,  00146 Roma, Italy  \\
~~E-mail: zullo@fis.uniroma3.it}

\end{center}

\medskip
\medskip

\begin{abstract}
\noindent
A two-parameters family of  B\"acklund transformations for the classical elliptic Gaudin model is constructed. The maps are explicit, symplectic, preserve the same integrals as for the continuous flows and are a time discretization of each of these flows. The transformations can map real variables into real variables, sending physical solutions of the equations of motion into physical solutions. The starting point of the analysis is the integrability structure of the model. It is shown how the analogue transformations for the rational and trigonometric Gaudin model are a limiting case of this one. An application to a particular case of the Clebsch system is given.     
\end{abstract}

\bigskip\bigskip\bigskip\bigskip

\noindent


\section{Introduction}
The Gaudin models describe completely integrable long range spin-spin systems, both at the classical and at the quantum level. They were first introduced by Gaudin \cite{Gaudin}, as the anisotropic, integrable generalization of what today is called the $xxx$, or isotropic, Gaudin model; the results of the analysis were the construction of the $xxz$ model, for the partially anisotropic case, and of the $xyz$ model, for the fully anisotropic case. The Lax matrices depend on the spectral parameter respectively through rational, trigonometric and elliptic functions. The Poisson structure of the models can be specified in the framework of the $r$-matrix approach \cite{Reyman}; in \cite{S1} Sklyanin showed how to obtain both the Lax matrices and the $r$-matrices of the models by a limiting procedure on the lattice Heisenberg magnet. Obviously the integrability structures of the trigonometric and rational cases can be also obtained as particular realizations of the corresponding structures of the more general elliptic model. From the point of view of separation of variables, functional Bethe ansatz and quantum inverse scattering method, a lots of results appeared in the late 80s-early 90s \cite{Jurco} \cite{KKM} \cite{Sk1} \cite{ST} \cite{ST1}. On the other hand a great number of works have pointed up until now very interesting connections between the Gaudin models and various branches of physics. For a recent list of the subjects involved the reader can see for example \cite{PS}. Here, in order to give an idea of the heterogeneity of the concerned issues, we wish only to mention the connections with the BCS theory and small metallic grains \cite{AFF}, with pairing models in nuclear physics \cite{DH}, with the Coulomb three-body problem \cite{Komarov}, with Lagrange and Kirchhoff tops \cite{MPR}. \newline 
In \cite{HKR}, \cite{SK1},\cite{SK2} the authors gave the B\"acklund transformations for the \emph{rational} Gaudin model in the framework of a larger research program started some years before by Kuznetsov and Sklyanin (see \cite{SK}) on the applications and properties of such transformation to finite dimensional integrable systems. In particular in \cite{HKR} it was shown how the two-parameters family of transformations obtained can be seen as a time discretizations of a family of flows and that the interpolating Hamiltonian flow encompasses all the commuting flows of the model. The features of the transformations found here are noteworthy: in fact the maps are (i) explicit, (ii) symplectic, (iii) they preserve all the integrals of the continuous model, (iv) they possess the so called \emph{spectrality} property \cite{KV} \cite{SK}. Furthermore they commute, that is the composition of two different transformations does not depend on the order of application, and with a two-parameters transformation it is possible to send real variable into real variables (so that the B\"acklund transformations send a real solution of the equations of motion into another real solution).   \newline
In the wake of these results in 2010 the analogue constructions for the trigonometric case were found \cite{noi} \cite{noi1}. All the above properties hold true in this case too. In \cite{noi1} the authors have also shown how, in the limit of small angles, the transformations obtained give exactly the results of the rational case as given in \cite{HKR}. \newline
The aim of this work is to complete the picture finding the B\"acklund transformations and the corresponding time-discretization for the elliptic Gaudin model. The paper is organized as follows: in section (\ref{sec1}) the main features and the integrability structures of the elliptic Gaudin model are briefly revised, in section (\ref{sec2}) the \emph{dressing} matrix is constructed; the parametrization obtained gives directly the dressing matrix of the trigonometric case \cite{noi1} when the elliptic modulus $k$ of the Jacobi elliptic function is set equal to zero. In section (\ref{sec3}) the \emph{explicit} form of the transformations is found by using the spectrality property. The symplecticity of the maps is also discussed. For any fixed set of initial conditions the transformations turn out to be birational maps. In section (\ref{sec4}) it is shown how it is possible to obtain a two-parameters family of physical transformations, that is transformations sending real variables into real variables. In section (\ref{sec5}) the continuous limit of the discrete dynamics is found: the interpolating Hamiltonian flow again encompasses all the commuting continuous flows of the model. Lastly, as an application of the results found, we construct the B\"acklund transformations for a particular case of the Clebsch system \cite{Clebsch} through a procedure of pole coalescence on the Lax matrix of the Gaudin model and taking advantage of the fact that this procedure preserves the $r$-matrix structure. The results are a generalization of the B\"acklund transformations for the Kirchhoff top as given in \cite{Kirchh}.   
\section{The elliptic Gaudin magnet} \label{sec1}
The elliptic Gaudin model is defined by the following Lax matrix:
\begin{equation}\label{eq:lax} 
L(\lm) =   \left( \begin{array}{cc} A(\lm) & B(\lm)\\C(\lm)&-A(\lm)\end{array} 
\right) 
\end{equation}
\begin{equation} 
\label{ABC}
A(\lm)=\sum_{j=1}^{N}\frac{\cnJ(\lm-\lm_{j})}{\snJ(\lm-\lm_{j})}s^{z}_{j}, \qquad 
B(\lm)=\sum_{j=1}^{N}\frac{s^{x}_{j}-\textrm{i}s^{y}_{j}\dnJ(\lm-\lm_{j})}{\snJ(\lm-\lm_{j})}
\end{equation}
\begin{equation*}
C(\lm)=\sum_{j=1}^{N}\frac{s^{x}_{j}+\textrm{i}s^{y}_{j}\dnJ(\lm-\lm_{j})}{\snJ(\lm-\lm_{j})}. 
\end{equation*} 
In (\ref{eq:lax}) and (\ref{ABC}) $\lm \in \mathbb{C}$ is the spectral parameter, $\lm_{j}$
are a set of $N$ arbitrary real parameters of the model, while
$\big(s^{x}_{j},s^{y}_{j},s^{z}_{j}\big)$,\, $j=1, \ldots, N$, are the
dynamical variables (spins) of the system.
In the \emph{r}-matrix formalism the Poisson structure is fixed by the
following equivalence \cite{ST}:
\begin{gather} \label{eq:pois} 
\big\{ L(\lm) , L(\mu)\big\}=\big[r_{e}(\lm-\mu), L(\lm)\otimes \dblone + \dblone\otimes L(\mu) \big], 
\end{gather} 
where $r_{e}$ is the elliptic solution to the classical Yang-Baxter
equation of Belavin and Drinfel'd \cite{FT},\cite{BD},\cite{ST}: 
\begin{gather} 
r_{e}(\lm) = \frac{\textrm{i}}{\snJ(\lm)}\left(\begin{array}{cccc} \cnJ(\lm)& 0 & 0 &\frac{1-\dnJ(\lm)}{2} \\ 
0 & 0 & \frac{1+\dnJ(\lm)}{2} & 0\\ 
0 & \frac{1+\dnJ(\lm)}{2} & 0 & 0\\ 
\frac{1-\dnJ(\lm)}{2} & 0 & 0 & \cnJ(\lm) 
\end{array}\right), 
\end{gather} 
The \emph{r}-matrix structure (\ref{eq:pois}) entails the following Poisson
brackets for the functions (\ref{ABC}):
\begin{equation}\label{ABCell}\begin{split} 
& \{A(\lm),A(\mu)\}=0,\\
&\{B(\lm),B(\mu)\}=\textrm{i}\,(A(\lm)+A(\mu))\frac{\dnJ(\lm-\mu)-1}{\snJ(\lm-\mu)}\\
&\{C(\lm),C(\mu)\}=\textrm{i}\,(A(\lm)+A(\mu))\frac{1-\dnJ(\lm-\mu)}{\snJ(\lm-\mu)}\\ 
& \{A(\lm),B(\mu)\}=\textrm{i}\,\frac{C(\lm)(1-\dnJ(\lm-\mu))-B(\lm)(1+\dnJ(\lm-\mu))+2B(\mu)\cnJ(\lm-\mu)}{2\snJ(\lm-\mu)}\\ 
& \{A(\lm),C(\mu)\}=\textrm{i}\,\frac{B(\lm)(\dnJ(\lm-\mu)-1)+C(\lm)(1+\dnJ(\lm-\mu))-2C(\mu)\cnJ(\lm-\mu)}{2\snJ(\lm-\mu)}\\ 
& \{B(\lm),C(\mu)\}=\textrm{i}\,\frac{(A(\mu)-A(\lm))(1+\dnJ(\lm-\mu))}{\sin(\lm-\mu)}.
\end{split}\end{equation}
Equivalently, the spin variables $\big(s^{x}_{j},s^{y}_{j},s^{z}_{j}\big)$,
$j=1..N$, have to obey to the corresponding algebra: 
\begin{gather} \label{poisS}
\big\{s^{x}_{j},s^{y}_{k}\big\}=\delta_{jk}s^{z}_{k}, \qquad  \big\{s^{y}_{j},s^{z}_{k}\big\}=\delta_{jk}s^{x}_{k},\qquad
\big\{s^{z}_{j},s^{x}_{k}\big\}=\delta_{jk}s^{y}_{k}
\end{gather} 
Due to the direct sum structure of the Poisson bracket (\ref{poisS}), the
square length of each spin is a Casimir function for the elliptic Gaudin
model, so we have $N$ Casimirs, given by  
\begin{equation*}
(s_{j}^{x})^{2}+(s_{j}^{y})^{2}+(s_{j}^{z})^{2}\doteq s_{j}^{2}\qquad j=1..N
\end{equation*}
Furthermore the model possesses $N$ integrals of motion in involutions
w.r.t. the Poisson brackets (\ref{poisS}): the determinant of the Lax matrix
is a generating function of such integrals (see \ref{Appendix A}): 
\begin{equation}\label{generfun}
-\textrm{det}(L)=A^{2}(\lambda)+B(\lambda)C(\lambda)=\sum_{i=1}^{N}\left(\frac{s_{i}^{2}}{\snJ^{2}(\lambda-\lambda_{i})}+2\varpi
 H_{i}\zeta(\varpi(\lambda-\lambda_{i}))\right)-H_{0}
\end{equation}
where $\zeta$ is the Weierstra\ss{} zeta function,
$\varpi=(e_{1}-e_{2})^{-\frac{1}{2}}$, $e_{i}=\wp(\frac{w_{i}}{2})$ and
$(w_{1},w_{2})$ are the periods of the Weierstra\ss{} $\wp$ function (see
\ref{Appendix A}). The $N$ Hamiltonians $H_{i}$ are given by:
\begin{equation} \label{hams}
H_{i}=\sum_{k\neq i}^{N} \frac{s_{i}^{z}s_{k}^{z}\cnJ(\lm_{i}-\lm_{k})+s_{i}^{y}s_{k}^{y}\dnJ(\lm_{i}-\lm_{k})+ s_{i}^{x}s_{k}^{x}}{\snJ(\lm_{i}-\lm_{k})} 
\end{equation}
Note that only $N-1$ among these Hamiltonians are independent, because of
$\sum_{i}H_{i}=0$. The other integral $H_{0}$ is given by the formula (see \ref{Appendix A}): 
\begin{equation} \begin{split}
H_{0}=&\sum_{i,k}^{N}(s_{i}^{z}s_{k}^{z}\dnJ(\lm_{i}-\lm_{k})+k^{2}s_{i}^{y}s_{k}^{y}\cnJ(\lm_{i}-\lm_{k}))+\\
+&\sum_{k\neq
i}^{N}\frac{a(\lm_{i}-\lm_{k})\big(s_{i}^{z}s_{k}^{z}\cnJ(\lm_{i}-\lm_{k})+s_{i}^{y}s_{k}^{y}\dnJ(\lm_{i}-\lm_{k})+ s_{i}^{x}s_{k}^{x}\big)}{\snJ(\lm_{i}-\lm_{k})},\\
&a(\lambda)\doteq \varpi\big(\zeta(\varpi\lm)-\zeta(2\varpi\lm)\big)-\frac{1}{\snJ(2\lm)} 
\end{split}\end{equation}  
Due to the existence of an $r$-matrix, the Hamiltonians $H_{i}$ are in involution for the Poisson bracket
(\ref{poisS}):
\begin{equation}
\{H_{i},H_{j}\}=0 \quad i,j=0,\ldots ,N-1
\end{equation}
The  corresponding Hamiltonian flows  are then given by: 
\begin{equation} \label{flows}
\frac{ds^{3}_{j}}{dt_{i}}=\{H_{i},s^{3}_{j}\}         \qquad  
\frac{ds^{\pm}_{j}}{dt_{i}}=\{H_{i},s^{\pm}_{j}\} 
\end{equation}
\section{The dressing matrix}\label{sec2}
By an Hamiltonian point of view, B\"acklund Transformations for
finite-dimensional integrable systems are (families of) symplectic maps
preserving the integrals of motion \cite{KV}: so we are searching for an
expression relating the dynamical variables
$\big(s^{x}_{j},s^{y}_{j},s^{z}_{j}\big)$ to the new set of variables $\big(\tl{s}^{x}_{j},\tl{s}^{y}_{j},\tl{s}^{z}_{j}\big)$       
and such that the brackets (\ref{poisS}) and the integrals (\ref{hams})
are preserved. The generating function of the integrals is given by
the determinant of the Lax matrix (\ref{eq:lax}), so the dressed Lax matrix
$\tl{L}(\lm)$, that is the Lax matrix of the tilded dynamical variables, has
to have the same determinant of $L(\lm)$. This means that the two Lax matrices are related by
a similarity transformation by means of a \emph{dressing} matrix $D(\lm)$:
\begin{equation}\label{simil}
\tilde{L}(\lm)D(\lm-\lm_{0})=D(\lm-\lm_{0})L(\lm)
\end{equation}
Here $\lm_{0}$ will be one of the two  B\"acklund parameters: actually we will
construct a parametric family of transformations and this will be a crucial
point when we will request to the maps to be ``physical'', that is to send
real variables in real variables.  
The rational and trigonometric Gaudin models are limiting case of the elliptic
one. The dressing matrices for the Lax matrices of these models are given, respectively, by
the elementary Lax matrix of the $xxx$ and $xxz$ Heisenberg spin chain on the
lattice \cite{HKR}, \cite{noi1}. It is obvious, for the dressing matrix of the
elliptic Gaudin model, to make
the ansatz of the elementary Lax matrix of the $xyz$ Heisenberg spin chain.
Note however that this matrix has to enjoy the same symmetry properties
with the Lax matrix (\ref{eq:lax}). In fact, the quasi-periodicity (\ref{quasip}) of the
Jacobi elliptic functions entails the following formulae for $L(\lm)$ (see also
\cite{ST1}):
\begin{equation} \label{symmL}
L(\lm+2K)=\sigma_{3}L(\lm)\sigma_{3}\qquad L(\lm+2iK')=\sigma_{1}L(\lm)\sigma_{1}
\end{equation}
where with $\sigma_{i}$, $i=1,2,3$ we indicate the Pauli matrices, and $K$ and
$K'$ are respectively the complete elliptic integral of the first kind and the
complementary integral (\ref{elliptic integrals}).
\newline
The above points suggest to make the following ansatz for $D(\lm)$
\begin{equation}\label{DdiF}
D(\lambda)=\mathscr{S}_{0}1+\frac{i}{\snJ(\lambda)}\big(\mathscr{S}_{1}\sigma_{1}+\dnJ(\lm)\mathscr{S}_{2}\sigma_{2}+\cnJ(\lambda)\mathscr{S}_{3}\sigma_{3}\big) \end{equation}
This is exactly the one-site Lax matrix for the $xyz$ Heisenberg spin chain on
the lattice \cite{FT}. The symmetries (\ref{symmL}) are also preserved: 
\begin{equation}\label{symmD}
D(\lm+2K)=\sigma_{3}D(\lm)\sigma_{3}\qquad D(\lm+2iK')=\sigma_{1}D(\lm)\sigma_{1}
\end{equation}
So far, $\mathscr{S}_{i}$, $i=0..3$ are four
undetermined variables, but we are free to fix one of them because of the
homogeneity of the equation (\ref{simil}).  The Lax matrix (\ref{eq:lax}) has
simple poles at the points $\lm=\lm_{j}$ (mod $2K,2iK'$), $\,j=1..N$; the
relation (\ref{simil}) is an equivalence between meromorphic functions (the
elements of the matrices), so that we have to equate the residue at the poles
on both sides. Because of the symmetries (\ref{symmL}) and (\ref{symmD}) we
can look only at the poles in $\lm=\lm_{j}$, $\,j=1..N$, that is: 
\begin{equation} \label{resk}
\tl{L}_{j}D_{j}=D_{j}L_{j} 
\end{equation} 
where 
\begin{equation}\label{Lk}  
L_{j}=\left(\begin{array}{cc}s^{z}_{j} & s^{x}_{j}-is^{y}_{j}\\ s^{x}_{j}+is^{y}_{j} & 
  -s^{z}_{j}\end{array}\right),\qquad D_{j}=D(\lambda=\lambda_{j}) 
\end{equation} 
In principle equation (\ref{resk}) gives an implicit relationship between the
old (untilded) variables and the new (tilded)
ones. It is however possible to get an explicit relationship by recurring to the so-called spectrality
property \cite{KV} \cite{SK}. To this aim, one need to force the determinant of
the Darboux matrix $D(\lm)$ to have two \emph{nondynamical} zeroes for two
arbitrary value of the spectral parameter $\lm$, say for $\lm=\lm_{0}\pm\mu$. This leaves us with only
\emph{two} undetermined variables in (\ref{DdiF}). As we will see, the
spectrality property will fixes these two variables, that we will call $P$ and $Q$,
as functions of only the untilded dynamical variables, so that the maps defined by (\ref{resk})
will be explicit. \\
Summarizing, by fixing for simplicity $\mathscr{S}_{0}=1$ and imposing the constraints
$$det(D(\lm-\lm_{0}))\Big|_{\lm=\lm_{0}\pm\mu}=0$$ 
we are left with two undetermined parameters,
that we denote with $P$ and $Q$: by choosing a particular parametrization of
the constraints we can write:
\begin{equation} \label{eq:Darboux} 
D(\lambda)=\left(\begin{array}{cc} 1+i\mathscr{S}_{3}\frac{\cnJ(\lm)}{\snJ(\lm)} & 
  \frac{i\mathscr{S}_{1}+\mathscr{S}_{2}\,\dnJ(\lm)}{\snJ(\lm)}\\ 
\frac{i\mathscr{S}_{1}-\mathscr{S}_{2}\,\dnJ(\lm)}{\snJ(\lm)} & 1-i\mathscr{S}_{3}\frac{\cnJ(\lm)}{\snJ(\lm)} 
\end{array}\right) \quad \textrm{with}\left\{\begin{aligned} &i\mathscr{S}_{3}=\frac{PQ-\snJ(\mu)}{\cnJ(\mu)}\\&i\mathscr{S}_{1}+\mathscr{S}_{2}\,\dnJ(\mu)=P\\&i\mathscr{S}_{1}-\mathscr{S}_{2}\,\dnJ(\mu)=Q(2\,\snJ(\mu)-PQ)\end{aligned}\right.
\end{equation} 
We recall again that $P$ and $Q$ are undetermined dynamical variable and that
$\lm_{0}$ and $\mu$ are constants: they are parameters for the B\"acklund
transformations. Note also that with this parametrization, in the limit $k\to 0$, one obtains the
dressing matrix for the trigonometric Gaudin model \cite{noi} (up to a trivial
multiplicative factor inessential for the form of the B\"acklund
transformation, as explained before).
\section{B\"acklund transformations}\label{sec3}
Now we make use of the spectrality property to find $P$ and $Q$ in terms of
only one set of variables, the untilded ones. The matrices $(D(\lm-\lm_{0}))\Big|_{\lm=\lm_{0}+\mu}$ and
$(D(\lm-\lm_{0}))\Big|_{\lm=\lm_{0}-\mu}$ are of rank one. We call
$|\Omega_{+}\rangle$ and $|\Omega_{-}\rangle$ their respective kernels. By
acting with these kernels on the equivalence defining the B\"acklund
transformations, we see that they are also the eigenvectors of
$L(\lm_{0}+\mu)$ and $L(\lm_{0}-\mu)$:
\begin{equation}\label{eigrel}
\tilde{L}(\lm_{0}\pm\mu)D(\pm \mu)|\Omega_{\pm}\rangle=0=D(\pm
\mu)\left[L(\lm_{0}\pm\mu)|\Omega_{\pm}\rangle\right]\Longrightarrow
L(\lm_{0}\pm\mu)|\Omega_{\pm}\rangle=\gamma_{\pm}|\Omega_{\pm}\rangle
\end{equation}
By viewing the generating function of the integrals (\ref{generfun}) as a
function of $\lm$, we define:
\begin{equation}\label{defgm}
\gm^2(\lm)\doteq-det(L(\lm))=A(\lm)^{2}+B(\lm)C(\lm)
\end{equation}
where $A(\lm)$, $B(\lm)$ and $C(\lm)$ are given by (\ref{ABC}). 
The two eigenvalues are then given by
$\gamma_{\pm}=\gamma(\lm)\Big|_{\lm=\lm_{0}\pm\mu}$. The two kernels
$|\Omega_{\pm}\rangle$ are expressions of the variables $P$ and $Q$, so that
the eigenvectors relations (\ref{eigrel}) for $L(\lm_{0}\pm\mu)$ link these
variables with the elements of the Lax matrix of the untilded
variables. Explicitly, the two kernels are given by: 
\begin{equation} 
|\Omega_{+}\rangle=\left(\begin{array}{c}1\\-Q\end{array}\right) \qquad |\Omega_{-}\rangle=\left(\begin{array}{c}P\\2\,\snJ(\mu)-PQ\end{array}\right) 
\end{equation} 
and these expressions in turns lead to the formulae:
\begin{equation}\label{eq:P&Q} 
Q=Q(\lambda_{0}+\mu)=\left.\frac{A(\lambda)-\gm(\lambda)}{B(\lambda)}\right|_{\lambda=\lambda_{0}+\mu} \qquad \frac{1}{P}=\frac{Q(\lambda_{0}+\mu)-Q(\lambda_{0}-\mu)}{2\,\snJ(\mu)} 
\end{equation} 
Note that, for arbitrary number $N$ of interacting spins of the model,  $P$
and $Q$ contain \emph{all} the dynamical variables so that the B\"acklund maps
touch all the spin sites. These maps associate to a given solution of the
equations of motion (\ref{flows}) a new solution. By fixing the initial
conditions, the generating function (\ref{generfun}), and therefore the
function $\gm(\lm)$, is a constant independent of time. As will be clear
in the next lines, if $\gm(\lm)$ is constant, then the B\"acklund transformations are
actually rational maps (or, better to say, birational maps). 
The equation (\ref{resk}) allow us to write the explicit transformations as follows:   
\begin{equation}\label{trans}\begin{split}
&\tl{s}^{x}_{k}=\frac{\left((\alpha^2_k+\varsigma^2_k-\beta^2_k-\delta^2_k)s^{x}_{k}+i(\delta^{2}_k+\varsigma^2_k-\beta^2_k-\alpha^2_k)s^{y}_{k}-2(\alpha_k\beta_k-\varsigma_k\delta_k)s^{z}_{k}\right)}{2\Delta_k}\\
&\tl{s}^{y}_{k}=\frac{\left(i(\alpha^2_k+\delta^2_k-\varsigma^2_k-\beta^2_k)s^{x}_{k}+(\beta^2_k+\varsigma^2_k+\alpha^2_k+\delta^2_k)s^{y}_{k}-2i(\beta_k\alpha_k+\varsigma_k\delta_k)s^{z}_{k}\right)}{2\Delta_k}\\
&\tl{s}^{z}_{k}=\frac{\left((\beta_k\varsigma_k-\alpha_k\delta_k)s^{x}_{k}+i(\beta_k\varsigma_k+\alpha_k\delta_k)s^{y}_{k}+(\alpha_k\varsigma_k+\beta_k\delta_k)s^{z}_{k}\right)}{\Delta_k}
\end{split}\end{equation}
where for brevity of notation we have introduced the functions $(\alpha_k,
\beta_k, \delta_k, \Delta_k, \varsigma_k)$ defined by the following formulae:
\begin{equation}\begin{split}
&\alpha_k=\snJ(\lm_k-\lm_{0})+\frac{PQ-\snJ(\mu)}{\cnJ(\mu)}\cnJ(\lm_k-\lm_{0})\\
&\beta_k=\frac{\left(P+Q(2\,\snJ(\mu)-PQ)\right)}{2}+\frac{\left(P-Q(2\,\snJ(\mu)-PQ)\right)}{2\,\dnJ(\mu)}\dnJ(\lm_k-\lm_0)\\
&\delta_k=\frac{\left(P+Q(2\,\snJ(\mu)-PQ)\right)}{2}-\frac{\left(P-Q(2\,\snJ(\mu)-PQ)\right)}{2\,\dnJ(\mu)}\dnJ(\lm_k-\lm_0)\\
&\varsigma_k=\snJ(\lm_k-\lm_{0})-\frac{PQ-\snJ(\mu)}{\cnJ(\mu)}\cnJ(\lm_k-\lm_{0})\\
&\Delta_k=\alpha_k\varsigma_k-\beta_k\delta_k
\end{split}\end{equation} 
When the elliptic modulus $k$ of the Jacobi elliptic functions is zero, the
transformations (\ref{trans}) coincide with those for the trigonometric Gaudin
magnet as given in \cite{noi}. At this point we have to deal with the
symplecticity of our maps. As the transformations are explicit, the
direct path to prove their symplecticity might be to use 
(\ref{poisS}) in  order to show that indeed the Poisson structure is
preserved. But this path is presumably not so plain because of the huge
calculations. As in \cite{noi}, we will follow a finer argument due to
Sklyanin \cite{SK2}.
Consider the relation (\ref{simil}) in an extended phase space, whose
coordinate are given by $(s^x_k, s^y_k, s^z_k, P, Q)$ and suppose that
$D(\lambda)$ obeys to the \emph{quadratic} Poisson bracket, as follows:
\begin{equation}\label{eq:PoisD} 
\{D(\lambda)\otimes\dblone,\dblone\otimes D(\tau)\}=[r_e(\lambda-\tau),D(\lambda)\otimes D(\tau)] 
\end{equation}
In the extended space we have to re-define (\ref{simil}) as:
\begin{equation}\label{eq:transf} 
\tilde{L}(\lambda)\tilde{D}(\lambda-\lambda_{0})=D(\lambda-\lambda_{0})L(\lambda) 
\end{equation} 
In fact in the left hand side of the previous one has to  use tilded
variables also for $D(\lambda)$ because (\ref{eq:transf}) defines  the
B\"acklund transformation  in the  extended phase space, where there is  also
a $\tl{P}$ and a $\tl{Q}$. 
Note that in the new phase space the entries of $D$ Poisson commutes with those of $L$. 
The key observation is that if both $L$ and $D$ have the 
same Poisson structure, given by equation (\ref{eq:PoisD}), then this property
holds true for $LD$ and 
$DL$ as well, because of the Poisson commutativity of the entries of $L$ and $D$. This means that the 
transformation (\ref{eq:transf}) defines a ``canonical'' 
transformation. Sklyanin showed \cite{SK2} that if one now restricts the variables on the constraint manifold $\tl{P}=P$ and $\tl{Q}=Q$ the symplecticity is 
preserved; however this constraint leads to a dependence of $P$ and $Q$ on 
the entries of $L$, that for consistency must be the same as the one given by the equation (\ref{eq:transf}) on this constrained manifold. But there 
(\ref{eq:transf}) reduce to (\ref{simil}),       
so that the map  preserves the spectrum of $L(\lambda)$ and is canonical. 
What remains to show is that indeed 
(\ref{eq:PoisD}) is fulfilled by our $D(\lambda)$.  
For the B\"acklund transformations of the rational Gaudin magnet the dressing
matrix has the quadratic Poisson structure imposed by the rational $r$-matrix
provided $P$ and $Q$ are canonically conjugated in the extended space
\cite{SK2}. In the trigonometric case one need to have a non trivial bracket
between $P$ and $Q$ in the extended space in order to guarantee the simplecticity of
B\"acklund transformations \cite{noi}. As we will show in the next lines, in
the elliptic case we found a non trivial bracket that, in the limit $k\to 0$
goes to the trigonometric result as given in \cite{noi}.   
By a direct inspection it is possible to show that (\ref{eq:PoisD}) entails
the following brackets between the elements $\mathscr{S}_i$, $i=0..3$ \cite{FT}:\begin{equation}\label{FTbra}
\begin{array}{cc} 
\{\mathscr{S}_{i},\mathscr{S}_{0}\}=J_{jk}\mathscr{S}_{j}\mathscr{S}_{k}\\ 
\{\mathscr{S}_{i},\mathscr{S}_{j}\}=-\mathscr{S}_{0}\mathscr{S}_{k} \end{array}
\end{equation}
where $(i,j,k)$ is a cyclic permutation of $(1,2,3)$ with  $J_{12}=k^2$,
$J_{23}=1-k^2$, $J_{31}=-1$. With the following positions: 
\begin{equation}\label{S0123}
\left\{\begin{aligned}&(\mathscr{S}_0)^2=\frac{\cnJ(\mu)\dnJ(\mu)}{\snJ(\mu)\left(1-2\,\snJ(\mu)PQ+P^2Q^2-k^2\Big[(PQ-\snJ(\mu))^2+\cnJ(\mu)^2\frac{(Q(PQ-2\,\snJ(\mu))-P)^2}{4}\Big]\right)}\\ &i\mathscr{S}_{3}=\frac{PQ-\snJ(\mu)}{\cnJ(\mu)}\mathscr{S}_0\\&i\mathscr{S}_{1}+\mathscr{S}_{2}\,\dnJ(\mu)=P\mathscr{S}_0\\&i\mathscr{S}_{1}-\mathscr{S}_{2}\,\dnJ(\mu)=Q(2\,\snJ(\mu)-PQ)\,\mathscr{S}_0\end{aligned}\right.
\end{equation}
after some calculations one can show that indeed (\ref{FTbra}) are fullfilled
provided that:
\begin{equation}\label{PQ}
\{Q,P\}=\textrm{i}\frac{\left(1-2\,\snJ(\mu)PQ+P^2Q^2-k^2\Big[(PQ-\snJ(\mu))^2+\cnJ(\mu)^2\frac{(Q(PQ-2\,\snJ(\mu))-P)^2}{4}\Big]\right)}{\cnJ(\mu)\dnJ(\mu)}
\end{equation} 
So the symplecticity of our maps is proved.
At this point let us to stress some remarks. The dressing matrix defined by
the relations (\ref{S0123}) is, except for the multiplicative factor
$\mathscr{S}_0$, completely equivalent to the dressing matrix as given by the
equation (\ref{eq:Darboux}). As explained before, given the homogeneity of the
equation (\ref{simil}), a factor of proportionality between two dressing
matrices is inessential as regards B\"acklund transformations, so by this
point of view the definitions (\ref{S0123}) are compatible with
(\ref{eq:Darboux}). Secondly, the Poisson bracket (\ref{PQ}) between $P$ and $Q$
reduces exactly, in the limit $k \to 0$, to the  bracket of the corresponding
variables in the trigonometric case. 
\section{Physical B\"acklund transformations}\label{sec4}
In this section we will show that with an appropriate choice of the parameters
$\lm_0$ and $\mu$ in (\ref{trans}), the B\"acklund transformations map real
solutions in real solutions, so in this sense the transformations can be
considered ``physical''. The choice amounts to require $\lm_{0}$ to be a real
number and $\mu$ to be a purely imaginary number. So, hereafter in this
section, we put:    
\begin{equation}\label{phys}
\mu=i\epsilon \qquad (\lm_0,\epsilon) \in \mathbb{R}^2
\end{equation}
The matrices $L_{j}$, $j=1..N$ defined in (\ref{Lk}) and corresponding to the real solutions $\big(s^{x}_{j},s^{y}_{j},s^{z}_{j}\big)$ of the
equations of motion are Hermitian. The request for physical B\"acklund
transformations is equivalent to the request for Hermitian dressed matrices
$\tl{L}_j$. By (\ref{resk}) we see that this means to have dressing matrices
$D_j$ proportional to unitary matrices. We claim that indeed when (\ref{phys})
are fulfilled then $D_{j}$ are of the form:
\begin{equation}\label{Dk}  
D_j =   \left( \begin{array}{cc} \alpha_j & \beta_j\\-\bar{\beta}_j&\bar{\alpha}_j\end{array} 
\right)
\end{equation}
where the bar  means complex conjugation. For clarity let us make the
following positions:
\begin{equation}
\lm_+=\lm_0+i\epsilon, \qquad \lm_{-}=\bar{\lm}_+
\end{equation}
We observe that, for the functions $A, B, C$, as defined in (\ref{ABC}),
one has:
\begin{equation}
A(\lm_{+})=\bar{A}(\lm_{-}), \qquad B(\lm_{+})=\bar{C}(\lm_{-}), \qquad C(\lm_{+})=\bar{B}(\lm_{-}).
\end{equation}
These relations entails $\gm^2(\lm_{+})=\bar{\gm}^2(\lm_{-})$. Note that this
last relation implies that the coefficients of the series of $\gm^2(\lm)$ with respect to $\lm$ are real, consistently with the expansion (\ref{generfun}). We recall that the
matrices $D_j$ are written in terms of $P$ and $Q$, that are defined by the
relations 
\begin{equation}
Q=Q(\lambda_{+})=\frac{A(\lambda_+)-\gm(\lambda_+)}{B(\lambda_+)}=-\frac{C(\lm_+)}{A(\lm_+)+B(\lm_+)} \qquad P=\frac{2\,\snJ(i\epsilon)}{Q(\lm_+)-Q(\lm_{-})}
\end{equation}
%
By specifying the sign of the function $\gm$ on the Riemann surface by $\gm(\lm_{+})=-\bar{\gm}(\lm_{-})$, one has:
$$
\bar{Q}(\lm_+)=-\frac{1}{Q(\lm_{-})}
$$
and this equation in turns implies that the matrices
$D_j=D(\lm)\Big|_{\lm=\lm_j}$, with $D(\lm)$ given by (\ref{eq:Darboux}), are
of the form (\ref{Dk}), with $\alpha_j$ and $\beta_j$ given by the following
formulae:
\begin{equation}\begin{split}
&\alpha_j=1+\frac{\snJ(i\epsilon)\,\cnJ(\lm_j)}{\cnJ(i\epsilon)\,\snJ(\lm_j)}\frac{|Q|^2-1}{|Q|^2+1}\\
&\beta_j=\frac{\snJ(i\epsilon)}{\snJ(\lm_j)}\left(\frac{\bar{Q}+Q}{|Q|^2+1}+\frac{\dnJ(\lm_j)}{\dnJ(i\epsilon)}\frac{\bar{Q}-Q}{|Q|^2+1}\right)
\end{split}\end{equation} 
So, under the given assumptions, the matrices $D_j$ are proportional to unitary matrices.
%
%
%
%
%
\section{Interpolating Hamiltonian flow}\label{sec5}
Now we want to get the interpolating flow of the discrete dynamics generated by the maps (\ref{trans}). As we will see the B\"acklund transformation can be seen as a time discretization of a
one-parameter ($\lambda_{0}$) family of Hamiltonian flows with the difference
$2\epsilon$ playing the role of the time-step and with the Hamiltonian defining the interpolating flow given by $\gm(\lm_0)$, where $\gm(\lm)$ is defined in (\ref{defgm}). \newline 
First of all let we take the limit $\epsilon \to 0$.\\
One has:
\begin{equation}\label{eeq:q0}
Q=\frac{A(\lambda_{0})-\gamma(\lambda_{0})}{B(\lambda_{0})}+O(\epsilon),
\end{equation}
\begin{equation}\label{eeq:p0}
P=-i\epsilon\frac{B(\lambda_{0})}{\gamma(\lambda_{0})}+O(\epsilon^{2}).
\end{equation}
One can carefully insert these expressions in the dressing matrix (\ref{eq:Darboux}) to find:
\begin{equation}\label{dto0}
D(\lambda-\lm_0)=\dblone-\frac{i\epsilon}{\gm(\lm_0)\snJ(\lm-\lm_0)}D_0(\lm, \lm_0),
\end{equation}
where
\be \label{d0}
D_0(\lm,\lm_0)\doteq\begin{pmatrix} A(\lm_0)\cnJ(\lm-\lm_0)& \frac{B(\lm_0)+C(\lm_0)}{2}+ \frac{B(\lm_0)-C(\lm_0)}{2}\dnJ(\lm-\lm_0)\\
\frac{B(\lm_0)+C(\lm_0)}{2}-\frac{B(\lm_0)-C(\lm_0)}{2}\dnJ(\lm-\lm_0)&- A(\lm_0)\cnJ(\lm-\lm_0)\end{pmatrix}.
\ee
In the  limit $\epsilon \to 0$ the equation of the map
$\tl{L}D=DL$ turns into the Lax equation for a continuous flow: 
\begin{equation}\label{eeq:motion}
\dot{L}(\lm)= [L(\lambda),M(\lambda,\lambda_{0})].
\end{equation}
where the time derivative is defined as:
\begin{equation}
\dot{L}=\lim_{\epsilon\rightarrow 0}\frac{\tilde{L}-L}{2\epsilon}
\end{equation}
and the matrix $M(\lambda,\lambda_{0})$ is given by:
\begin{equation}
M(\lm,\lm_0)=\frac{i}{2\gm(\lm_0)\snJ(\lm-\lm_0)}D_0(\lm,\lm_0).
\end{equation}
With the help of the Poisson brackets between the elements of the Lax matrix (\ref{ABCell}), the dynamical system (\ref{eeq:motion})
can be cast in Hamiltonian  form:
\begin{equation}
\dot{L}_{ij}(\lambda)=\{\mathcal{H}(\lambda_{0}),L_{ij}(\lambda)\}, \qquad i,j\in \{1,2\},
\end{equation} 
with the Hamilton's function given by:
\begin{equation}\label{eeq:Ham}
\mathcal{H}(\lambda_{0})=\gamma(\lambda_{0})=\sqrt{A^{2}(\lambda_{0})+B(\lambda_{0})C(\lambda_{0})}.
\end{equation}
So the Hamiltonian (\ref{eeq:Ham})
characterizing the interpolating flow is (the square root of) the generating
function (\ref{generfun}) of the whole set of conserved quantities. By
choosing the parameter $\lambda_{0}$ to be equal to any of the poles ($\lambda_{i}$) of the
Lax matrix, the map leads to $N$ different maps $\{BT^{(i)}\}_{i=1..N}$, where
$BT^{(i)}$ discretizes the flow
corresponding to the Hamiltonian $H_{i}$, given by equation
(\ref{hams}). In fact, by posing $\lambda_{0}=\delta +\lambda_{i}$ and taking
the limit $\delta \to 0$, the Hamilton's function (\ref{eeq:Ham}) gives:
\begin{equation}
\gamma(\lambda_{0})=\frac{s_{i}}{\delta}+\frac{H_{i}}{s_{i}}+O(\delta).
\end{equation}
and the equations of motion take the form:
\begin{equation}
\dot{L}_{ij}(\lambda)=\frac{1}{s_{i}}\{H_{i},L_{ij}(\lambda)\}, \qquad i,j\in \{1,2\}.
\end{equation}
Note that the corresponding interpolating Hamiltonian flows of the B\"acklund transformations for the trigonometric and rational Gaudin models can be obtained as the corresponding limiting case of this one.
%
%
%
%
%
\section{An application to the Clebsch model}\label{sec6}
The Clebsch model \cite{Clebsch} is an integrable case of the Kirchhoff equations
\cite{Kirchhoff} describing the motion of a solid in an infinite incompressible
fluid. \newline 
If the solid has three perpendicular planes of symmetry and there are no external forces then the Kirchhoff system can be described in terms of an Hamiltonian (the kinetic energy of the system solid+fluid) quadratic and diagonal in the \emph{impulsive force} $\boldsymbol{p}$ and \emph{impulsive pair} $\boldsymbol{J}$ vectors, representing respectively the sum of the impulse
and angular momentum of the solid and those applied by the solid to the boundary of the fluid in contact with it \cite{Milne}. \newline
The Hamiltonian then reads: 
\be\label{CS}
T=\frac{1}{2}\left(\alpha_{1}(p^x)^2+\alpha_{2}(p^y)^2+\alpha_{3}(p^z)^2\right)+\frac{1}{2}\left(\beta_{1}(J^x)^2+\beta_{2}(J^y)^2+\beta_{3}(J^z)^2\right),
\ee
where $\alpha_{i}$ and $\beta_{i}$ are a set of constants depending on the shape of the solid. Clebsch \cite{Clebsch} discovered that if the following constraint on these quantities holds:
\be\label{Cleconst}
\frac{\alpha_{1}-\alpha_{2}}{\beta_{3}}+\frac{\alpha_{2}-\alpha_{3}}{\beta_{1}}+\frac{\alpha_{3}-\alpha_{1}}{\beta_{2}}=0,
\ee
then the corresponding equations of motion are integrable. \newline
Since the dynamical variables are impulse and angular momentum vectors, the Lie-Poisson structure is defined by the \emph{e}(3) algebra:
\begin{equation}\label{Poisson}
\{J^{i},J^{j}\}=\epsilon^{ijk}J^{k}, \quad
\{J^{i},p^{j}\}=\epsilon^{ijk}p^{k}, \quad \{p^{i},p^{j}\}=0.
\end{equation}
where $i,j,k$ belong to the set $\{x,y,z\}$. These brackets have two Casimirs:
\begin{equation}\label{Casimir}
\boldsymbol{p}\cdot\boldsymbol{J}\doteq c_{1}, \qquad \boldsymbol{p}^{2}\doteq c_{2}. 
\end{equation}
Now we will show how, by the means of a procedure of \emph{pole coalescence} on the Lax
matrix of the two-site elliptic Gaudin model, it is possible to obtain the Lax matrix for the Clebsch model governed by the following Hamiltonian:
\be\label{MyH}\begin{split}
2\mathcal{H}=&\left(C-k^2(A+B)\right)(p^x)^2+C(p^y)^2+\left(C+B(1-k^2)\right)(p^z)^2+B(J^x)^2+\\
&+\left(A k^2+B\right)(J^y)^2+\left(A+B\right)(J^z)^2.
\end{split}\ee
The construction of the integrability structure of the model is not new (cf. with \cite{PR}), but here is briefly reported for completeness. For more details about the pole coalescence procedure see \cite{MPR}, \cite{PR}. \newline  
Note that the Clebsch's constraint (\ref{Cleconst}) holds for (\ref{MyH}), but here we have only four (not five) arbitrary constants, so actually we will obtain a special realization of (\ref{CS}). Note also that in the case $k=0$ one obtains the Kirchhoff top \cite{Kirchh}. \newline
First let us introduce a contraction parameter, say
$\epsilon$, and take in the Lax matrix (\ref{eq:lax}) $\lambda_{1}\to\epsilon
\lambda_{1}$ and $\lambda_{2}\to\epsilon\lambda_{2}$, where we recall that $\lambda_1$ and $\lambda_2$ are the two arbitrary parameters of the Gaudin model.
By setting:
\begin{equation}\label{Jp}
\mathbf{J}\doteq \mathbf{s}_{1}+\mathbf{s}_{2}, \qquad \mathbf{p}\doteq
\epsilon (\lm_{1}\mathbf{s}_{1}+\lm_{2}\mathbf{s}_{2})
\end{equation}
and letting $\epsilon \to 0$ in (\ref{eq:lax}) {\it{after}} this
identification, one obtains the following expression:
\begin{equation}\label{eq:laxC} \begin{split}
&\snJ(\lm)L(\lm)=\\  
&=\left(\begin{array}{cc}
  \cnJ(\lm)J^{z}+p^{z}\frac{\dnJ(\lm)}{\snJ(\lm)}&J^{x}-\textrm{i}\,\dnJ(\lm)J^{y}+\frac{\cnJ(\lm)\dnJ(\lm)}{\snJ(\lm)}p^x-\textrm{i}\frac{\cnJ(\lm)}{\snJ(\lm)}p^y\\
J^{x}+\textrm{i}\,\dnJ(\lm)J^{y}+\frac{\cnJ(\lm)\dnJ(\lm)}{\snJ(\lm)}p^x+\textrm{i}\frac{\cnJ(\lm)}{\snJ(\lm)}p^y&-\cnJ(\lm)J^{z}-p^{z}\frac{\dnJ(\lm)}{\snJ(\lm)}\end{array}\right)  
\end{split}\end{equation}
The overall term $\snJ(\lm)$ multiplying $L(\lm)$ on the l.h.s. of the previous equation can be obviously skipped, so in the following we assume that the Lax matrix of the model. that we will call $L^c$, is given only by the r.h.s. term of (\ref{eq:laxC}).
Note that by using (\ref{poisS}), it is readily seen that the
variables $\mathbf{J}$ and $\mathbf{p}$ (\ref{Jp}) obey the Lie-Poisson algebra \emph{e}(3) (\ref{Poisson}).\newline
The determinant of this matrix is the generating function of the integrals of
motions. One has:
\begin{equation}\label{genfun}
-det(L^c(\lm))\doteq\Upsilon^2(\lm) = H_{1}-H_{0}\,\snJ^2(\lm)+2c_{1}\frac{\cnJ(\lm)\dnJ(\lm)}{\snJ(\lm)}+c_{2}\frac{\cnJ^2(\lm)}{\snJ^2(\lm)}
\end{equation}
where $c_{1}$ and $c_{2}$ are the Casimirs (\ref{Casimir}), while $H_{0}$ and
$H_{1}$ are the two Poisson commuting integrals:
\begin{equation}\label{integrals}
H_{0}=(J^z)^2+k^2\left((J^y)^2-(p^x)^2\right), \quad H_{1}=\mathbf{J}^2+(p^z)^2-k^2\left((p^x)^2+(p^z)^2\right).
\end{equation}
The physical hamiltonian (\ref{MyH}) is now obtained by the following linear combination:
$$
2\mathcal{H}\doteq A\,H_0+B\,H_1+C\,c_1
$$
\subsection{B\"acklund transformations} The ansatz for the dressing matrix for the model just described is inherited from that of the elliptic Gaudin model since the r-matrix structure is preserved by the pole coalescence
procedure. So for the general form of $D(\lm)$ we again refer to (\ref{eq:Darboux}):
\begin{equation} \label{eq:DarbouxC} 
D(\lambda)=\left(\begin{array}{cc} 1+i\mathscr{S}_{3}\frac{\cnJ(\lm)}{\snJ(\lm)} & 
  \frac{i\mathscr{S}_{1}+\mathscr{S}_{2}\,\dnJ(\lm)}{\snJ(\lm)}\\ 
\frac{i\mathscr{S}_{1}-\mathscr{S}_{2}\,\dnJ(\lm)}{\snJ(\lm)} & 1-i\mathscr{S}_{3}\frac{\cnJ(\lm)}{\snJ(\lm)} 
\end{array}\right) \quad \textrm{with}\left\{\begin{aligned} &i\mathscr{S}_{3}=\frac{pq-\snJ(\mu)}{\cnJ(\mu)}\\&i\mathscr{S}_{1}+\mathscr{S}_{2}\,\dnJ(\mu)=p\\&i\mathscr{S}_{1}-\mathscr{S}_{2}\,\dnJ(\mu)=q(2\,\snJ(\mu)-pq)\end{aligned}\right.
\end{equation}
Exactly as for the Gaudin model, the variables $p$ and $q$ can be determined in terms of only one set of variables thanks to the spectrality property. The result is:
\begin{equation}\label{eq:pq} 
q=q(\lambda_{0}+\mu)=\left.\frac{L^{c}_{11}(\lambda)-\Upsilon(\lambda)}{L^{c}_{12}(\lambda)}\right|_{\lambda=\lambda_{0}+\mu} \qquad \frac{1}{p}=\frac{q(\lambda_{0}+\mu)-q(\lambda_{0}-\mu)}{2\,\snJ(\mu)} 
\end{equation} 
Again by choosing $\lm_0$ real and $\mu$ purely imaginary one obtains transformations sending real variables into real variables. To write explicitly the maps one can, for example, take the residue at the pole in $\lm=0$ in the equivalence $\tl{L}^c(\lm)D(\lm-\lm_0)=D(\lm-\lm_0)L^c(\lm)$ and its value at $\lm=K(k)$, where $K(k)$ is the complete elliptic integral of the first kind (\ref{elliptic integrals}). We write directly the real transformations by posing:
\begin{equation}
\mu=i\epsilon \qquad (\lm_0,\epsilon) \in \mathbb{R}^2.
\end{equation}
They reads:
\begin{equation}\label{pJtilde}\begin{split}
&\tl{p}^x=\frac{a^2_0+\bar{a}^2_0-b^2_0-\bar{b}^2_0}{2 h_0}p^x+\textrm{i}\frac{\bar{a}^2_0-a^2_0+\bar{b}^2_0-b^2_0}{2 h_0}p^y-\frac{a_0b_0+\bar{a}_0\bar{b}_0}{h_0}p^z\\
&\tl{p}^y=\frac{a^2_0+\bar{a}^2_0+b^2_0+\bar{b}^2_0}{2 h_0}p^y+\textrm{i}\frac{a^2_0-\bar{a}^2_0+\bar{b}^2_0-b^2_0}{2 h_0}p^x-\textrm{i}\frac{a_0b_0-\bar{a}_0\bar{b}_0}{h_0}p^z\\
&\tl{p}^z=\frac{|a_0|^2-|b_0|^2}{h_0}p^z+\frac{a_0\bar{b}_0+b_0\bar{a}_0}{ h_0}p^x+\textrm{i}\frac{b_0\bar{a}_0-a_0\bar{b}_0}{h_0}p^y\\
&\tl{J}^x=\frac{a^2_1+\bar{a}^2_1-b^2_1-\bar{b}^2_1}{2 h_1}J^x+\textrm{i}k'\,\frac{\bar{a}^2_1-a^2_1-\bar{b}^2_1+b^2_1}{2 h_1}J^y-k'\,\frac{a_1b_1+\bar{a}_1\bar{b}_1}{h_1}p^z\\
&\tl{J}^y=\frac{a^2_1+\bar{a}^2_1+b^2_1+\bar{b}^2_1}{2 h_1}J^y+\textrm{i}\frac{a^2_1-\bar{a}^2_1+\bar{b}^2_1-b^2_1}{2\,k'\, h_1}J^x-\textrm{i}\frac{a_1b_1-\bar{a}_1\bar{b}_1}{h_1}p^z
\end{split}\end{equation}
where the functions $(a_i,
b_i, h_i)$, $i\in\{0,1\}$, are defined by the following formulae:
\begin{equation}\label{abh}\begin{split}
&a_0\doteq \snJ(i\epsilon)\cnJ(\lm_0)\left(|q|^2-1\right)-\snJ(\lm_0)\left(|q|^2+1\right)\\
&b_0\doteq \snJ(i\epsilon)\left(q+\bar{q}\right)-\frac{\snJ(i\epsilon)}{\dnJ(i\epsilon)}\dnJ(\lm_0)\left(q-\bar{q}\right)\\
&a_1\doteq \frac{\cnJ(\lm_0)}{\dnJ(\lm_0)}\left(|q|^2+1\right)+k'\,\frac{\snJ(i\epsilon)\snJ(\lm_0)}{\cnJ(i\epsilon)\cnJ(\lm_0)}\left(|q|^2-1\right)\\
&b_1\doteq \snJ(i\epsilon)\left(q+\bar{q}\right)-k'\,\frac{\snJ(i\epsilon)}{\dnJ(i\epsilon)\dnJ(\lm_0)}\left(q-\bar{q}\right)\\
&h_i\doteq |a_i|^2+|b_i|^2\qquad i\in\{0,1\}
\end{split}\end{equation}
In the previous formulae $k'$ is the complementary modulus of the Jacobi elliptic functions, $k'=\sqrt{1-k^2}$; the bar means complex conjugation. The expression for $\tl{J}^z$ follows for example by the constraint $\tl{\mathbf{J}}\cdot\tl{\mathbf{p}}=\mathbf{J}\cdot\mathbf{p}$, but a cleaner expression can be found taking the value $\lm=K(k)+\textrm{i}K'(k)$ in $\tl{L}^c(\lm)D(\lm-\lm_0)=D(\lm-\lm_0)L^c(\lm)$, where $K'(k)$ is the complementary integral (\ref{elliptic integrals}). It reads:
\begin{equation} \begin{split}  
&\tl{J}^z=\frac{a_2^2-b_2^2-a_3^2+b_3^2}{h}J^z-2\,k\,\frac{a_3b_2+a_2b_3}{ h}J^1-2\,k\,\frac{a_2a_3+b_2b_3}{h}p^y\\
&\textrm{where}\quad \left\{\begin{aligned} &a_2\doteq\frac{\dnJ(\lm_0)}{\cnJ(\lm_0)}\left(|q|^2+1\right)\\& b_2\doteq\textrm{i}k'\frac{\snJ(\textrm{i}\epsilon)}{\cnJ(\textrm{i}\epsilon)\cnJ(\lm_0)}\left(|q|^2-1\right)\end{aligned}\right.\quad
\left\{\begin{aligned} 
&a_3\doteq -\textrm{i}k\,\snJ(\textrm{i}\epsilon)\left(q+\bar{q}\right)\\& b_3\doteq k\,k'\,\frac{\snJ(\textrm{i}\epsilon)\snJ(\lm_0)}{\dnJ(\textrm{i}\epsilon)\cnJ(\lm_0)}\left(q-\bar{q}\right)\\& h\doteq a_2^2+a_3^2-b_2^2-b_3^2 
\end{aligned}\right.
\end{split}\end{equation}
Note that if in these transformations one poses $k=0$, then the B\"acklund transformations for the Kirchhoff top as given in \cite{Kirchh} are obtained. \newline
Now let us consider the interpolating Hamiltonian flow. We recall that the $r$-matrix structure of the model is that of Gaudin, so the Poisson brackets (\ref{ABCell}) works again by substituting $A=L^c_{11}$, $B=L^c_{12}$ and $C=L^c_{21}$. This is enough to ensure that the interpolating Hamiltonian is given, as for the Gaudin models, by the square root of the generating function of the integrals of the system (\ref{genfun}) for $\lm=\lm_0$:
\begin{equation}
\mathcal{H}(\lambda_{0})=\Upsilon(\lambda_{0})=\sqrt{H_{1}-H_{0}\,\snJ^2(\lm_0)+2c_{1}\frac{\cnJ(\lm_0)\dnJ(\lm_0)}{\snJ(\lm_0)}+c_{2}\frac{\cnJ^2(\lm_0)}{\snJ^2(\lm_0)}}.
\end{equation} 
With a choice of the parameter $\lm_0$ it is possible to obtain a discretization of the continuous flow corresponding to each linear combination of the Hamiltonians $H_0$ and $H_1$. \newline
A last remark: the symplecticity of these maps simply follows from the symplecticity of the maps for the ancestor Gaudin model, again thanks to the preservation of the $r$-matrix structure. 
%
%
%
%
%
\section{Comments}
First of all we have to mention the lacks of our construction. Unlike the rational case (cf. with \cite{HKR}), we are not able to give the generating function of the canonical transformations defined by the maps (\ref{trans}). However this isn't only a matter of technical difficulties; indeed for the $xxx$ Gaudin model the two-parameters B¨acklund transformations can be written as the composition of two simpler one-parameter transformations:
the same property holds true for the generating functions; yet in the trigonometric case a factorization of the dressing matrix
cannot lead to a one parameter dressing matrix preserving all the symmetries of the problem (\cite{noi1}). There are two possibilities: or one is able to find directly the two parameters generating function, or one should look for a symmetry-violating generating function such that their composition restores the symmetries. These details can be interesting by a quantum point of view for their potential connections with the Baxter's Q operator \cite{SBaxter}. \newline
In \cite{Kirchh} it was shown how the discrete orbits defined by the B\"acklund transformations for the Kirchhoff top exactly interpolate, in some special cases and in general as a conjecture, the continuous orbits of the corresponding physical flow, indicating that indeed the B\"acklund transformations can be considered an integrator (numerical or analytical) of the corresponding continuous differential problem. It could be interesting to understand how many of these results remain true for the transformations here given, both for Gaudin and Clebsch models. Works are in progress in this direction.

%
%
%
%
%
\appendix
\section{Notations and formulae}\label{Appendix A}
Let $w_{1}, w_{2}$ be complex numbers such that their ratio is not real and
consider the lattice $\Lambda$ generated by these numbers:
\begin{equation*}
\Lambda=\{w\in \mathbb{C}:w=n_{1}w_{1}+n_{2}w_{2}, n_{1},n_{2}\in\mathbb{Z}^{2}\}
\end{equation*}
The Weierstra\ss{} zeta function is given by \cite{Armitage}:
\begin{equation}
\zeta(u)=\frac{1}{u}+\sum_{w\neq 0}^{}\left(\frac{1}{u-w}+\frac{1}{w}+\frac{u}{w^{2}}\right)
\end{equation}
The Weierstra\ss{} $\wp$ function is minus the derivative of $\zeta$:
\begin{equation}
\wp(u)=-\zeta'(u)=\frac{1}{u^{2}}+\sum_{w\neq 0}\left(\frac{1}{(u-w)^{2}}-\frac{1}{w^{2}}\right)
\end{equation}
By denoting the period $w_{3}$ such that $w_{1}+w_{2}+w_{3}=0$ and defining
the set $e_{i},\, i=1..3$ by $e_{i}=\wp(\frac{w_{i}}{2})$, then
 holds the relation \cite{Armitage}:
\begin{equation}\label{wpsn}
\wp(u\varpi)=e_{2}+\frac{1}{\varpi^{2}\snJ(u,k)}
\end{equation}
where $\varpi=(e_{1}-e_{2})^{-\frac{1}{2}}$ and the elliptic modulus for the
  Jacobi ``sn'' function is given by $k^{2}=\varpi^{2}(e_{3}-e_{2})$.
The Jacobi elliptic functions $\snJ(u,k)$, $\cnJ(u,k)$ and $\dnJ(u,k)$ satisfies the
  following quasi-periodic relations \cite{Armitage}:
\begin{equation}\label{quasip}\begin{split}
&\snJ(u+2mK+2inK',k)=(-1)^{m}\snJ(u,k)\\
&\cnJ(u+2mK+2inK',k)=(-1)^{m+n}\cnJ(u,k)\\
&\dnJ(u+2mK+2inK',k)=(-1)^{n}\dnJ(u,k)
\end{split}\end{equation}
where $K$ and $K'$ are respectively the complete elliptic integral of the first kind and the
complementary integral:
\begin{equation}\label{elliptic integrals}
K(k)=\int_{0}^{1}\frac{dt}{\sqrt{(1-t^{2})(1-k^{2}t^{2})}}\qquad K'(k)=\int_{0}^{1}\frac{dt}{\sqrt{(1-t^{2})(1-(1-k^{2})t^{2})}}
\end{equation}
The following formulae are useful in proving (\ref{generfun}):
\begin{equation}\label{cnxy}
\cnJ(x\pm y)=\cnJ(x)\cnJ(y)\mp\dnJ(x\pm y)\snJ(x)\snJ(y)
\end{equation}
\begin{equation}\label{dnxy}
\dnJ(x\pm y)=\dnJ(x)\dnJ(y)\mp k^{2}\cnJ(x\pm y)\snJ(x)\snJ(y)
\end{equation}
\begin{equation}\label{funceq}\left\{\begin{array}{ll}
&\varpi\left(\zeta(\varpi x)-\zeta(\varpi
y)\right)-\frac{\snJ(y-x)}{\snJ(x)\snJ(y)}= a(y-x)\\
&a(x)\doteq\varpi\left(\zeta(\varpi (x))-\zeta(2\varpi x)\right)-\frac{1}{\snJ(2x)}
\end{array}\right.\end{equation}
Equations (\ref{cnxy}) and (\ref{dnxy}) are only a consequence of addition
formulae for the Jacobi elliptic functions, (\ref{funceq}) can be proved in
few lines. In fact suppose that $x$ and $y$ vary while $y-x$ remains constant
and equal to $b$. Differentiating $f(x)=\varpi\left(\zeta(\varpi x)-\zeta(\varpi
(x+b))\right)-\frac{\snJ(b)}{\snJ(x)\snJ(x+b)}$ with respect to $x$ we see
that this function is independent of $x$. In fact
$$
f'(x)=\varpi^{2}\big(\wp((x+b)\varpi)-\wp(x\varpi)\big)-\frac{\snJ(b)\big(\snJ(x)\snJ(x+b)\big)'}{\big(\snJ(x)\snJ(x+b)\big)^{2}}
$$   
By using the relation (\ref{wpsn}) and again the addition formulas for the
Jacobi elliptic functions it is readily shown that $f'(x)=0$, so $f(x)$ is a
constant, that we can take as a function of $b$. This implies the relation $\varpi\left(\zeta(\varpi x)-\zeta(\varpi
y)\right)-\frac{\snJ(y-x)}{\snJ(x)\snJ(y)}= a(y-x)$. By posing $y=2x$ in this
equation we obtain the function $a(x)$ as in (\ref{funceq}).
Now let us consider closely the formula for $-det(L(\lambda))$. For brevity we
pose in the following $v_{i}=\lambda-\lambda_{i}$ and $v_{ij}=\lambda_{i}-\lambda_{j}$. From (\ref{ABC}) we have:
\begin{equation}\begin{split}
-det(L(\lambda))=&\sum_{i,j}\frac{\cnJ(v_{i})\cnJ(v_{j})s^{z}_{i}s^{z}_{j}+s^{x}_{i}s^{x}_{j}+\dnJ(v_{i})\dnJ(v_{j})s^{y}_{i}s^{y}_{j}}{\snJ(v_{i})\snJ(v_{j})}=\\
&=
\sum_{\overset{i,j}{i\neq
  j}}\frac{\cnJ(v_{i})\cnJ(v_{j})s^{z}_{i}s^{z}_{j}+s^{x}_{i}s^{x}_{j}+\dnJ(v_{i})\dnJ(v_{j})s^{y}_{i}s^{y}_{j}}{\snJ(v_{i})\snJ(v_{j})}+\\
&+\sum_{i}\frac{s_{i}^{2}}{\snJ(v_{i})^{2}}-\sum_{i}\left((s_{i}^{z})^{2}+k^{2}(s_{i}^{y})^{2}\right)
\end{split}\end{equation}
By adding and subtracting the quantities $\sum_{i \neq j}
\left(\dnJ(v_{ij})s^{z}_{i}s^{z}_{j}+k^{2}\cnJ(v_{ij})s^{y}_{i}s^{y}_{j}\right)
$ in the last equation and using (\ref{cnxy}) and (\ref{dnxy}), we find:
\begin{equation}\label{interm}\begin{split}
-det(L(\lambda))=&\sum_{i}\frac{s_{i}^{2}}{\snJ(v_{i})^{2}}-\sum_{i,j}\left(s_{i}^{z}s_{j}^{z}\dnJ(v_{ij})+k^{2}s_{i}^{y}s_{j}^{y}\cnJ(v_{ij})\right)+\\
&+\sum_{\overset{i,j}{i\neq
  j}}\frac{\cnJ(v_{ij})s^{z}_{i}s^{z}_{j}+s^{x}_{i}s^{x}_{j}+\dnJ(v_{ij})s^{y}_{i}s^{y}_{j}}{\snJ(v_{i})\snJ(v_{j})}
\end{split}\end{equation}
Now, using formula (\ref{funceq}) on the denominator of the last sum of
equation (\ref{interm}) and defining $$H_{i}=\sum_{j\neq i}^{N}
\frac{s_{i}^{z}s_{j}^{z}\cnJ(v_{ij})+s_{i}^{y}s_{k}^{y}\dnJ(v_{ij})+
  s_{i}^{x}s_{k}^{x}}{\snJ(v_{ij})}$$ one reaches the result (\ref{generfun}).

\end{document}